\documentclass[sigconf]{acmart}

\AtBeginDocument{%
	\providecommand\BibTeX{{%
			Bib\TeX}}}

\usepackage{amsmath,amsfonts, amsthm}
\usepackage{booktabs} %
\usepackage{hyperref}
\usepackage[10pt]{moresize}
\usepackage{graphicx}
\usepackage[english]{babel}
\usepackage{dsfont}
\usepackage{latexsym}
\usepackage{csquotes}
\usepackage{tcolorbox}
\usepackage{listings}
\usepackage{float}
\usepackage{multirow}
\usepackage[scaled]{helvet}
\usepackage[noend]{algpseudocode}
\usepackage{mathrsfs}
\usepackage[linesnumbered, ruled, lined]{algorithm2e}  %
\usepackage{mathpartir}
\usepackage{mathtools}
\usepackage{dsfont} 
\usepackage{stmaryrd}
\usepackage{url}
\usepackage{textcomp} 
\usepackage{bbm}
\usepackage{verbdef}
\usepackage{xspace}
\usepackage{verbatim}
\usepackage{lipsum}
\usepackage{wrapfig}

\usepackage[shortlabels]{enumitem}
\usepackage{pifont}
\usepackage{varwidth}
\usepackage{xpatch}
\usepackage{multirow}
\usepackage{xcolor}
\usepackage[export]{adjustbox}
\usepackage{caption}
\usepackage{subcaption}
\usepackage{setspace}
\usepackage[capitalize]{cleveref}
\usepackage{makecell}
\usepackage{dashrule}
\usepackage{comment}
\usepackage{arydshln}
\usepackage{ifsym}
\usepackage{mdframed}
\usepackage{todonotes}

\usepackage[varqu]{zi4}

\usepackage{flushend}

\usepackage{tikz}
\usetikzlibrary{calc,positioning,math, arrows, shapes}
\newtheorem{definition}{Definition}

\AtBeginDocument{%
  \providecommand\BibTeX{{%
    \normalfont B\kern-0.5em{\scshape i\kern-0.25em b}\kern-0.8em\TeX}}}

\setlist[itemize]{topsep=0pt,itemsep=-1ex,partopsep=1ex,parsep=1ex,itemindent=0pt,leftmargin=8pt}
\newcommand{\toolname}{\emph{Conflux}}

\definecolor[named]{ACMBlue}{cmyk}{1,0.1,0,0.1}
\definecolor[named]{ACMYellow}{cmyk}{0,0.16,1,0}
\definecolor[named]{ACMOrange}{cmyk}{0,0.42,1,0.01}
\definecolor[named]{ACMRed}{cmyk}{0,0.90,0.86,0}
\definecolor[named]{ACMLightBlue}{cmyk}{0.49,0.01,0,0}
\definecolor[named]{ACMGreen}{cmyk}{0.20,0,1,0.19}
\definecolor[named]{ACMPurple}{cmyk}{0.55,1,0,0.15}
\definecolor[named]{ACMDarkBlue}{cmyk}{1,0.58,0,0.21}

\makeatletter
\def\arcr{\@arraycr}
\makeatother

\makeatletter
\newcommand{\mtmathitem}{%
\xpatchcmd{\item}{\@inmatherr\item}{\relax\ifmmode$\fi}{}{\errmessage{Patching of \noexpand\item failed}}
\xapptocmd{\@item}{$}{}{\errmessage{appending to \noexpand\@item failed}}}
\makeatother

\definecolor{shadecolor}{gray}{1.00}
\definecolor{ddarkgray}{gray}{0.75}
\definecolor{darkgray}{gray}{0.30}
\definecolor{light-gray}{gray}{0.87}

\newcommand{\etal}{\emph{et~al.}\xspace}

\definecolor{pblue}{rgb}{0.13,0.13,1}
\definecolor{pgreen}{rgb}{0,0.5,0}
\definecolor{pred}{rgb}{0.9,0,0}
\definecolor{pgrey}{rgb}{0.46,0.45,0.48}

\definecolor{ckeyword}{HTML}{7F0055}
\definecolor{ccomment}{HTML}{3F7F5F}
\definecolor{cnumber}{HTML}{2A0099}
\definecolor{darkgreen}{HTML}{008000}

\definecolor{clr-background}{RGB}{255,255,255}
\definecolor{clr-text}{RGB}{0,0,0}
\definecolor{clr-string}{RGB}{163,21,21}
\definecolor{clr-namespace}{RGB}{0,0,0}
\definecolor{clr-preprocessor}{RGB}{128,128,128}
\definecolor{clr-keyword}{RGB}{0,0,255}
\definecolor{clr-type}{RGB}{43,145,175}
\definecolor{clr-variable}{RGB}{0,0,0}
\definecolor{clr-constant}{RGB}{111,0,138} %
\definecolor{clr-comment}{RGB}{0,128,0}

\lstdefinestyle{VS2017}{
	backgroundcolor=\color{clr-background},
	basicstyle=\color{clr-text}, %
	stringstyle=\color{clr-string},
	identifierstyle=\color{clr-variable}, %
	commentstyle=\color{clr-comment},
	directivestyle=\color{clr-preprocessor}, %
	keywordstyle=\color{clr-type},
	keywordstyle={[2]\color{clr-constant}}, %
  tabsize=2,
}
\newcommand{\mathsc}[1]{{\normalfont\textsc{#1}}}

\newcommand{\callgraph}{G}
\usepackage{lipsum}

\usepackage{soul}

\usepackage{url}

\author{Jyoti Prakash}
\email{prakash@fim.uni-passau.de}
\affiliation{%
 \institution{University of Passau}
 \city{Passau}
 \country{Germany}
}

\author{Abhishek Tiwari}
\email{tiwari@fim.uni-passau.de}
\affiliation{%
 \institution{University of Passau}
 \city{Passau}
 \country{Germany}
}

\author{Christian Hammer}
\email{hammer@fim.uni-passau.de}
\affiliation{%
 \institution{University of Passau}
 \city{Passau}
 \country{Germany}
}

\definecolor{lightgray}{rgb}{.9,.9,.9}
\definecolor{darkgray}{rgb}{.4,.4,.4}
\definecolor{purple}{rgb}{0.65, 0.12, 0.82}
\lstdefinelanguage{JavaScript}{
  keywords={break, case, catch, continue, debugger, default, delete, do, else, false, finally, for, function, if, in, instanceof, new, null, return, switch, this, throw, true, try, typeof, var, void, while, with},
  morecomment=[l]{//},
  morecomment=[s]{/*}{*/},
  morestring=[b]',
  morestring=[b]",
  ndkeywords={class, export, boolean, throw, implements, import, this},
  keywordstyle=\color{blue}\bfseries,
  ndkeywordstyle=\color{darkgray}\bfseries,
  identifierstyle=\color{black},
  commentstyle=\color{purple}\ttfamily,
  stringstyle=\color{red}\ttfamily,
  sensitive=true
}

\begin{document}

\title{Unifying Pointer Analyses for Polyglot Inter-operations through Summary Specialization}

	\sloppy

\begin{abstract}
Modular analysis of polyglot applications is challenging because heap object flows across language boundaries must be resolved. The state-of-the-art analyses for polyglot applications have two fundamental limitations. First, they assume explicit boundaries between the host and the guest language to determine inter-language dataflows. Second, they rely on specific analyses of the host and guest languages. The former assumption is impractical concerning recent advancements in polyglot programming techniques, while the latter disregards advances in pointer analysis of the underlying languages. In this work, we propose to extend existing pointer analyses with a novel summary specialization technique so that points-to set across language boundaries can be unified. Our novel technique leverages various combinations of host and guest analyses with minor modifications. We demonstrate the efficacy and generalizability of our approach by evaluating it with two polyglot language models: Java-C communication via Android's NDK and Java-Python communication in GraalVM.

\end{abstract}

\maketitle

\section{Introduction}
Developers are rapidly adopting the polyglot programming model. Recent advancements in runtime systems such as GraalVM, paired with innovations in programming languages and models, have accelerated the adoption of the polyglot paradigm model~\cite{davidlo2016polyglot, mlpolyglotstudy, stackexchange, polyglotstudy2, rcpp, rpy, adlib2019ryu, hybridDroid, LuDroid-Conference, JuCify2022ICSE, hydridDroidICSE2019}. Undoubtedly, the polyglot programming model is frequently advantageous to developers. They can reuse or repackage existing modules written in any language into new applications. Unfortunately, %
the potential for misuse increases simultaneously: With a limited understanding of language interoperability, developers may introduce major flaws or even vulnerabilities%
~\cite{davidlo2016polyglot, mlpolyglotstudy, polyglotstudy2}. Consequently, it has become critical to develop tools and practices that facilitate polyglot development.

Static program analysis supports development of such tools. A modular technique, in which each language module is studied independently and the findings combined for a whole-program analysis, appears to be promising for scalability and precision~\cite{CousoutModular2002}. 
Recent years have seen a surge in research towards modular program analysis, with a multitude of new techniques being developed for popular languages~\cite{nielsen2021issta,schubert2021ecoop}. However, a modular analysis for polyglot applications is still in its infancy. A modular approach to inter-language analysis can benefit from a language's models and optimizations~\cite{Ryu-Semantic-Summary}. However, numerous obstacles avert this goal.

Integrating analysis results of several language models in a modular style is one of the main obstacles. Existing approaches, such as Lee~\etal~\cite{Ryu-Semantic-Summary} translate data-flow summaries of the guest language's operations to the host language. After injecting these summaries into the host program, this program is analyzed. However, there are two primary disadvantages to this approach. First, due to the complexity of modern languages' syntax and semantics, translating all conceivable consequences of the guest language to the host language is a complex and potentially error-prone undertaking. It entails modeling every language aspect of the guest language and transforming them into the host language's semantics. Lee~\etal~\cite{Ryu-Semantic-Summary} consequently only %
convert a \emph{subset} of the guest language's semantics to the host. Second, the translated program is analyzed using the host language's model and thus cannot take advantage of the guest language's unique characteristics, preventing language-specific optimizations for a most appropriate result of the guest module. %

Another drawback of present techniques is that they rely on explicit host-guest language boundaries~\cite{Monat-Python-SAS}. These boundaries are identified via explicit syntactic markers, which are typically predefined function calls, signatures, or data structures involved in interlanguage communication. Static analyses can detect these predefined syntactic structures and define models that replicate their behavior in the host language. However, current polyglot programming environments such as \emph{GraalVM} or Android's \emph{WebView} allow code from both languages to be mixed, blurring these explicit boundaries. For instance, WebViews allows messages from the guest to the host as well, which are not linked to explicit syntactic markers but communicate via a shared object. Prior research demonstrated that these functionalities are widespread in contemporary applications~\cite{LuDroid-Journal}. Furthermore, it is unavoidable that future multilingual programming environments will offer similar patterns, making it impractical to rely entirely on static language boundaries.

\paragraph*{Our Approach} We propose a novel strategy to address these challenges. Our modular technique is oblivious of the particular language models: It can instantiate various combinations of host and guest analyses to take advantage of the language-specific models, thereby inheriting the precision, scalability, and soundness improvements of the underlying analyses. Notably, our technique extends existing host and guest language analyses to include essential information concerning call and data flows to their counterparts. Given a polyglot model, our analysis finds the communication endpoints and features (shared classes/functions) of a host that a guest can use. With this information, an \emph{interlanguage call graph} is created by joining the missing call flow edges in the host's and guest's individual call graphs. Next, our technique builds an intra-procedural function summary of all the interoperating functions discovered in the interlanguage call graph. The function summaries are determined using intra-procedural points-to-analysis according to an \emph{Andersen style}~\cite{SridharanAndersen} constraint scheme. Next, these summaries are resolved such that the points-to sets of variables shared across languages contain the correct and complete references. To this end we propose innovative \emph{summary specialization} and \emph{unification} techniques. Unification is an alias-aware technique that unifies the variables' must-aliases across host and guest. The resolved summaries are then translated to be consistent with the host and guest's analysis models. With the resolved call edges and unified points-to set, the analysis is performed as a whole by merging the host and guest results independently.

\begin{figure}
	\begin{adjustbox}{width=\columnwidth}
    $\begin{array}{lll}
        m \in \mathit{host~methods}       & := & m_1 \mid m_2 \mid m_3 \mid \cdots \mid m_n \\
        i \in \mathit{InterfaceVars}  & :=  & i_1 \mid \cdots  \mid i_n \\
        v \in \mathit{variables}       & := & v_1 \mid v_2 \mid v_3 \mid \cdots \mid v_n \mid \mathit{InterfaceVars} \\
        g \in \mathit{fields} & :=  & g_1 \mid g_2 \mid g_3 \mid \cdots \mid g_n \\
        s \in \mathit{statements}       & := & v=\mathbf{new}~T()  \mid i=\mathbf{new}~T() \mid  \\
                                        & &  v' = v.m() \mid v' = v.g \mid v'.g = v  \mid \mathbf{if}~(e)~\{ s \} [\mathbf{else}~\{s\}]\\
                                        & &  v=\mathbf{eval}(program) \mid \mathbf{op}~s_1~s_2 \mid s_1;s_2 \mid \mathbf{while}~(e)~\{s\} \\
    \end{array}$
\end{adjustbox}
    \centerline{\textbf{Host Language}}
    \hdashrule{\columnwidth}{0.5pt}{1mm}
    \begin{adjustbox}{width=\columnwidth}
    $\begin{array}{lll}
        f \in \mathit{guest~methods}& := & f_1 \mid f_2 \mid f_3 \mid \cdots \mid f_n \\
        w \in \mathit{variables}& := & v_1 \mid v_2 \mid v_3 \mid \cdots \mid v_n \mid \mathit{InterfaceVars} \\
        h \in \mathit{fields} & := & h_1 \mid h_2 \mid h_3 \mid \cdots \mid h_n \\
        t \in \mathit{statements}       & ::=  & v=\mathbf{new}~T() \mid w' = w.f() \mid \\
                                        & &  w' = i.m() \mid w' = w.h \mid \mathbf{if}~(e)~\{ t \} [\mathbf{else}~\{t\}] \\ 
                                        &  &  w'.h = w \mid \mathbf{op}~t_1~t_2 \mid t_1;t_2 \mid \mathbf{while}~(e)~\{t\} \\
    \end{array}$
\end{adjustbox}
    \centerline{\textbf{Guest Language}}
    \caption{Host and guest language model}
    \label{fig:language-model}
\end{figure}

We implemented our approach into a tool called~\toolname. To evaluate the efficacy and generalizability of our approach, we applied \toolname~to two polyglot language models; 1) Java-C communication via the Android NDK~\cite{androidndk} and 2) Java-Python communication in GraalVM~\cite{graalvm}. \toolname~outperforms the previous state-of-the-art and successfully analyzed 31 out of 39 apps using Java-C communication in Android NDK. Besides, \toolname~successfully analyzed two of three Java-Python apps using GraalVM APIs. To the best of our knowledge, \toolname~is the first polyglot analysis tool that can be applied to multiple multilingual language models.

To summarize, this paper makes the following contributions:
\begin{enumerate}[(i), topsep=0em]
    \item \emph{Modular analysis using summary specialization.} We present a novel modular analysis framework to analyze multiple polyglot language models. Our framework can be employed on combinations of host and guest analyses to take advantage of the language-specific models.  
    \item \emph{Polyglot analysis tool.} We implement our approach into a tool called \toolname. \toolname~supports two widely used polyglot models, and it can be easily extended to analyze additional multilingual language models. \toolname~is publicly available~\cite{conflux_2022} for experiments and extensions.
    \item \emph{Extensive evaluation.} We evaluate \toolname~on multiple apps from two polyglot language models and demonstrate its efficacy and generalizability.
\end{enumerate}

\section{Modular Inter-language Analysis}

\paragraph*{A core polyglot model}
There are various polyglot language models, e.g. Java-C, Java-Javascript, Java-python, Javascript-wasm, Java-Javascript-wasm, and communication settings. 
In this work, we propose the first modular analysis framework for handling multiple polyglot language models. To this end, we define a language model sufficiently abstracted from the general polyglot model's semantics to ease the presentation and explanation of our approach.  Our language model, depicted in \cref{fig:language-model}, incorporates the semantics required for host-to-guest and guest-to-host communication in a polyglot model following the foreign function interface (FFI) model\footnote{Other potential communication facilities, such as middleware-based remote method invocation, have been covered by previous research (\cref{sec:relw}) and are thus out of scope of this approach.}. This language is based on an in-depth examination of existing polyglot programming environments, specifically GraalVM and Android hybrid apps. The host and guest languages have been defined as standard im\-per\-a\-tive/OO programs with the addition of inter-language communication constructs. Standard details such as class and field syntax are omitted to concentrate on the analysis. %
\begin{lstlisting}[caption={Host language code}, label={lst:toyexample:host}, language={JavaScript}, basicstyle=\footnotesize, frame=single, numbersep=1em, float, floatplacement=H, numbers=left]
(*\label{lstlistingA:1}*)foo() {
(*\label{lstlistingA:2}*)  i = new BridgeClass(); // obj@iface
(*\label{lstlistingA:3}*)  eval("set()"); // invoke the guest code
  // do some computation
(*\label{lstlistingA:4}*)  x = eval("getAndLeak()");
(*\label{lstlistingA:5}*)  return x;
}
(*\label{lstlistingA:6}*)BridgeClass.setSecret(secret) {
(*\label{lstlistingA:7}*)   this.secret = secret;  
}
(*\label{lstlistingA:8}*)BridgeClass.getSecret() {
(*\label{lstlistingA:9}*)   return this.secret;
}
\end{lstlisting}

\begin{lstlisting}[caption={Guest language example}, label={lst:toyexample:guest}, language={JavaScript}, basicstyle=\footnotesize, firstnumber=15, frame=single, numbersep=1em, float, floatplacement=H, numbers=left]
(*\label{lstlistingA:10}*)set() {
(*\label{lstlistingA:11}*)  v = new Secret(); //obj@sec 
(*\label{lstlistingA:12}*)  i.setSecret(v);
}
(*\label{lstlistingA:13}*)getAndLeak() {
(*\label{lstlistingA:14}*)  g = i.getSecret();
(*\label{lstlistingA:15}*)  leak(g);
(*\label{lstlistingA:16}*)  return g;
}
\end{lstlisting}

The functions, variables, field variables, and statements in both the host and guest languages are equivalent. 
 \emph{Interface variables} (\emph{InterfaceVars}) and \emph{eval} are the two key features that make it possible for the host and guest to communicate with each other. The host language construct \emph{eval} executes a guest language program or expression and returns the corresponding result to the host. The host language defines \emph{InterfaceVars} for the guest to access host functionalities. InterfaceVars are host objects that the guest can send messages to. Technically, they are ordinary objects that serve as a bridge between a host's class and the guest. Polyglot environments that allow a subset of the host language features to be accessed by the guest are a specific instance of this premise. \cref{lst:toyexample:host} and \cref{lst:toyexample:guest} employ our language constructs to exemplify a simplified polyglot model. \cref{lstlistingA:2} declares an interface variable \emph{i} of the host class \emph{BridgeClass}, granting the guest language access to its methods. In Lines~\ref{lstlistingA:3} and \ref{lstlistingA:4}, the host uses the \emph{eval} method to invoke guest language code. In turn at \cref{lstlistingA:12} and \ref{lstlistingA:14} in the guest, the \emph{BridgeClass}'s \emph{setSecret} and \emph{getSecret} functions are accessed.

\paragraph*{Analysis Overview}
\begin{figure}[tb]
	\includegraphics[height=6cm]{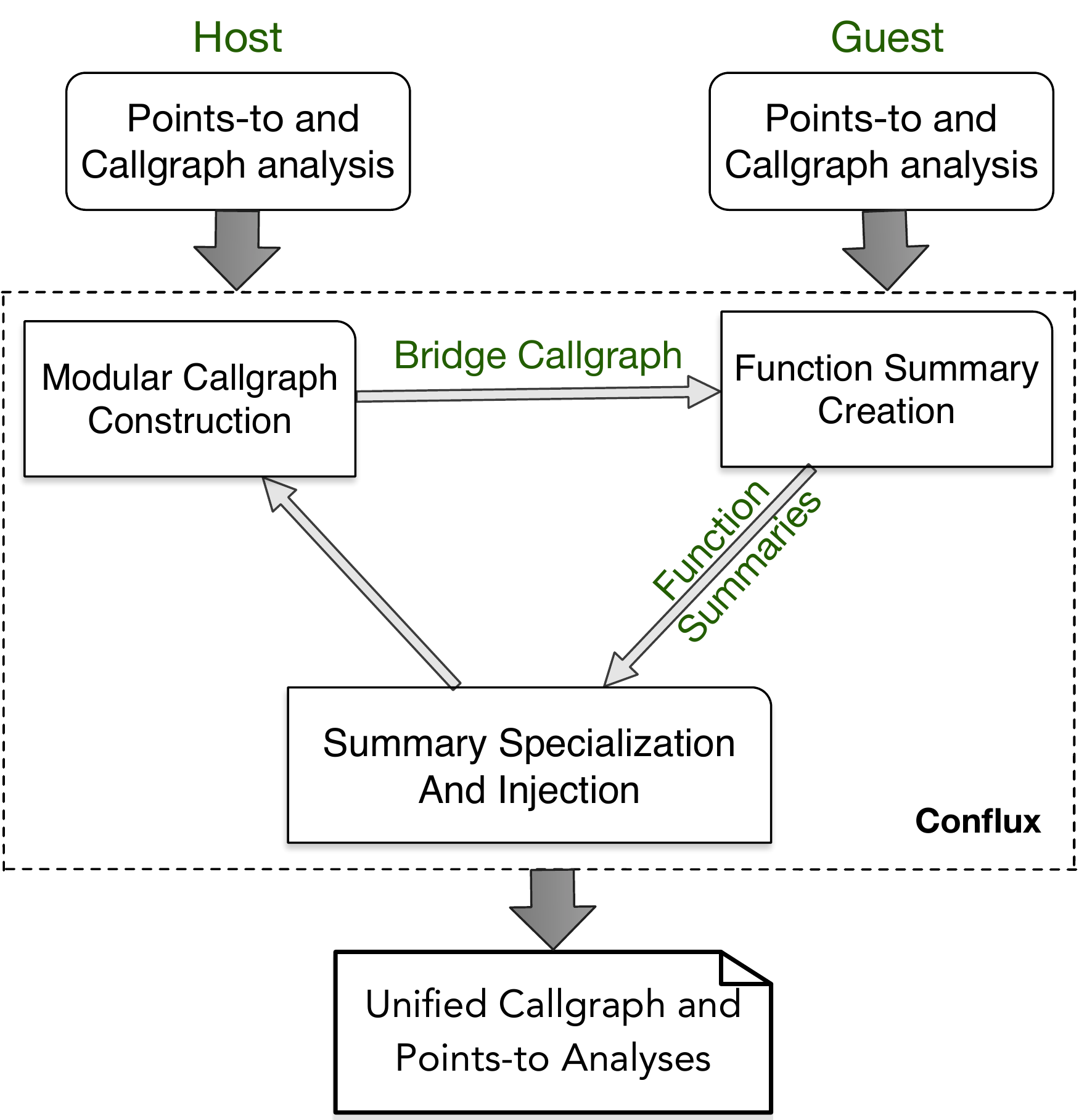}
	\caption{Analysis Overview}
	\label{fig:analysis-pipeline}
\end{figure}

\cref{fig:analysis-pipeline} depicts the components of our analysis based on the aforementioned language constructs. The framework builds upon a scalable context-sensitive pre-analysis for the host and guest languages. It then computes other building blocks of the analysis, a call graph (constructed with a modular analysis, \cref{sec:modular-call-graph-construction}) and intra-procedural function summaries (\cref{sec:function-summary-constraints}).
Next we elaborate on our bottom-up analysis strategy. First, we visualize our analysis for bridge calls initiated within a single method~(\cref{sec:pre-analysis}--\cref{sec:summary-injection}). Then, we extend our approach to model bridge function calls in multiple methods~(\cref{sec:multiple-bridge-calls}).

\subsection{Pre-analysis}\label{sec:pre-analysis}

In our approach, the \emph{pre-analysis} module leverages points-to and callgraph analyses computed independently for the host and guest languages (i.e., modulo the foreign function interfaces). This module provides flexibility in establishing the sensitivity levels of the underlining analysis for individual languages. As a result, it may adopt the most precise and scalable analyses specific to the particular languages. Next, we describe the pre-analysis model and exemplify it for the example in \cref{lst:toyexample:host} and \ref{lst:toyexample:guest}.

The pre-analysis is depicted in \cref{fig:pre-analyis-model}. It includes a set of variables $V$, a set of heap objects $H$, and a collection of contexts $C$. $C$ can be modeled in various notions; by using objects as contexts (object sensitive), by invoking function call sites (call-site sensitive), and many others. Context set $C$ improves the analysis by removing unrealizable paths (such as paths from unmatched function calls). A context $\mathit{ctxt}$ is a string that is constructed using elements from $C$. For the sake of presentation we may choose to omit contexts in the sequel.

\begin{figure}[tb]
\begin{tabular}{ll}
    $V : \text{Set of variables}$ & $C : \text{Set of contexts}$ \\
    $H : \text{Set of heap objects}$  &  $\mathit{Ctxt}^j  \subseteq c_1^j \times c_2^j \times \ldots \times c_n^j,~c_i^j \in C$ \\ 
\end{tabular}
\vspace{-1em}
\caption{Pre-analysis model}
\label{fig:pre-analyis-model}
\end{figure}

A points-to analysis generates a mapping from variables to a set of heap objects, referred here as \textit{variable points-to set} (under a context $\mathit{ctxt}$). A variable points-to mapping, $\mathit{VarPointsTo}: V \mapsto \mathcal{P}(H)$, is a mapping from the set of variables to heap objects. A points-to set for a variable $v \in V$ is a set of heap objects $h \subseteq H$. In the presence of fields in a programming language, one adds one more resolution by mapping object and field pairs to sets of heap objects, referred to as \textit{field points-to set}. If $f$ is a field for a heap object $o_i$, its points-to set is the set of heap objects, $O_\mathit{if}$, referred through the field $f$ of the heap object $o_i$. A field points-to mapping, $\mathit{FldPointsTo}: H \times \mathit{Field} \mapsto \mathcal{P}(H)$ maps each pair of a heap object and a field to a set of heap objects.

\begin{figure}
  	\begin{adjustbox}{width=0.89\linewidth}
        \begin{subfigure}{0.45\linewidth}
            \begin{tabular}{|l|}
                \hline
                $\langle i, ctx \rangle$  $\mapsto$  $\{\texttt{obj@iface}\}$ \\ 
                $\langle x, ctx \rangle$  $\mapsto$  $\emptyset$ \\ \hdashline[3pt/1pt] %
                $\langle g, ctx' \rangle$  $\mapsto$  $\emptyset$ \\
                $\langle v, ctx' \rangle$  $\mapsto$  $\{\texttt{obj@sec}\}$ \\ \hline
            \end{tabular}
            \caption{Points-to analysis}
             \label{fig:preanalysisA}
        \end{subfigure}
        \begin{subfigure}{0.45\linewidth}
        \begin{tabular}{|l|}
            \hline
            $\langle \mathit{foo}, \mathit{ctx} \rangle \rightarrow \langle \mathit{bar}, \mathit{ctx} \rangle $ \\ \hdashline[3pt/1pt] %
            $\langle \mathit{set}, \mathit{ctx'} \rangle$  $\rightarrow$  $\langle \mathit{setSecret}, \mathit{ctx'} \rangle$ \\
            $\langle \mathit{get}, \mathit{ctx'} \rangle$  $\rightarrow$  $\langle \mathit{getSecret}, , \mathit{ctx'} \rangle$ \\
            $\langle \mathit{get}, \mathit{ctx'} \rangle$  $\rightarrow$  $\langle \mathit{leak}, , \mathit{ctx'} \rangle$ \\ \hline %
        \end{tabular}
        \caption{Callgraph analysis}
    \end{subfigure}
\end{adjustbox}
\vspace{-1em}
    \caption{Pre-analysis of host and guest language}
    \label{fig:preanalysis}
\end{figure}

In the example in \cref{lst:toyexample:host} and \ref{lst:toyexample:guest}, a pre-analysis of the host language infers that $i$ and $v$ under some contexts %
refer to \texttt{obj@iface} and \texttt{obj@sec}, respectively (\cref{fig:preanalysisA}). Since $x$ captures an FFI call's return value, the analysis ignores this operation. Similarly, in the guest language, $g$ captures the value of an FFI call, which is also ignored. As expected, the pre-analysis result is incomplete and misses crucial data flows. \cref{lst:toyexample:host} and \ref{lst:toyexample:guest} show that the methods $\texttt{set}$ and $\texttt{get}$ are reachable from $\texttt{foo}$. \cref{sec:modular-call-graph-construction} extends the pre-analysis to resolve these edges by constructing an inter-language callgraph. Further, the guest object $\texttt{obj@sec}$ defined at \cref{lstlistingA:11} is written to the BridgeClass's field \texttt{secret} through the bridge invocation at \cref{lstlistingA:12}. At \cref{lstlistingA:14}, it reads from this field into $g$ and at the end returns it to the variable $x$ in \texttt{foo}. Therefore, $g$ and $x$ should refer to $\texttt{obj@sec}$. We resolve these dataflows with our modular callgraph construction and summary specialization approach (\cref{sec:summary-injection}).

\subsection{Modular Call-Graph Construction}\label{sec:modular-call-graph-construction}

The modular call-graph construction determines the edges for inter-language function calls happening through \emph{eval} and interface variables. It builds upon the incomplete call-graphs obtained from the pre-analysis phase. The process begins with an incomplete host call graph and expands it to include all bridge class methods; our analysis assumes that all bridge class methods can be invoked from the guest. Initially, the extended host call-graph ($\callgraph_{host}$) includes potentially disconnected nodes (we include methods of the bridge class even if the host code does not invoke them). The next objective is to complete the missing edges in the host and guest call graphs ($ \callgraph_{guest}$), i.e., to add the corresponding edges from host to guest (via \emph{eval}) and vice versa (via \emph{interface variables}).

\begin{figure*}[tb]
        \centering
        \includegraphics[scale=0.20]{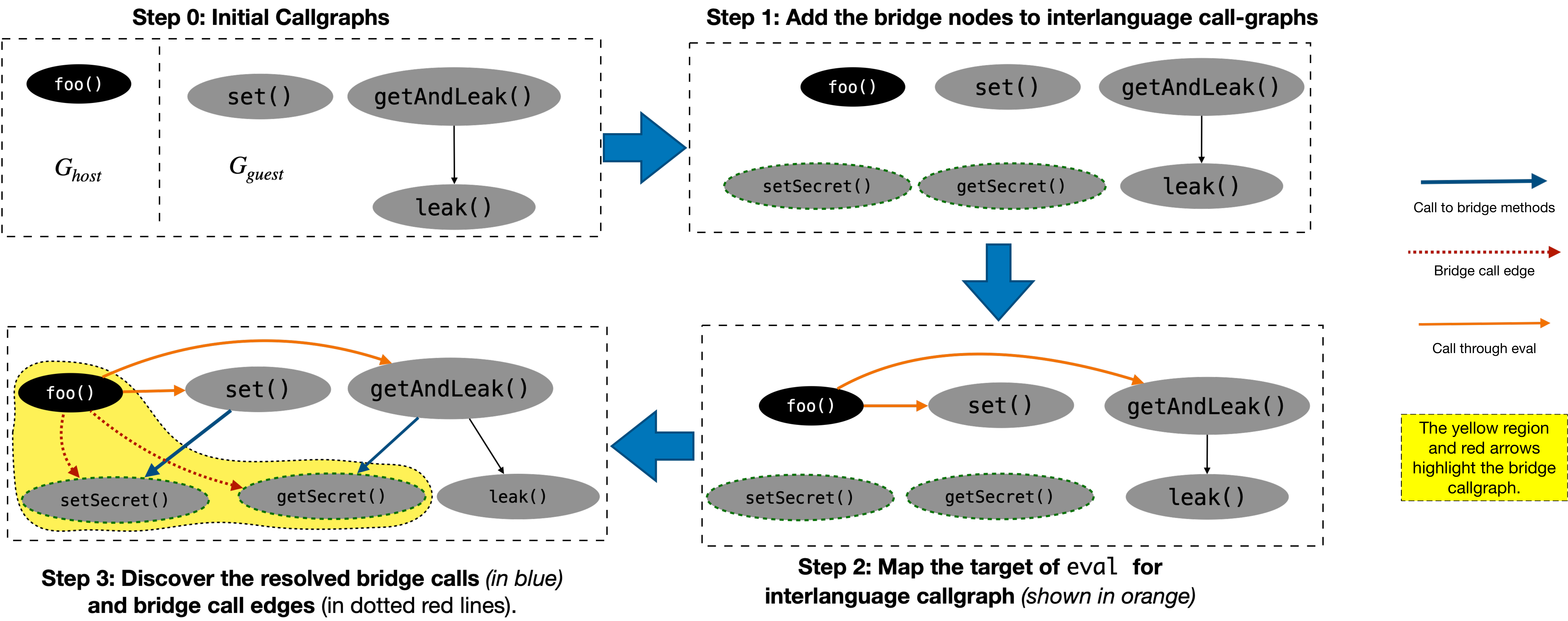}
        \caption{Call-graph for program in \cref{lst:toyexample:host} and \cref{lst:toyexample:guest}} 
        \label{fig:call-graph-example}
\end{figure*}

\begin{algorithm}[tb]
    \caption{Modular Callgraph Analysis}\label{algo:callgraph}
    
   \KwIn{Host and guest call graph $\callgraph_{host}$, $\callgraph_{guest}$ }
   \KwOut{Interlanguage callgraph $G$}
   \KwOut{Set of bridge callgraphs $B_{m_h}, m_h \in \callgraph_{host}$}
    $G \leftarrow \callgraph_{host} \cup \callgraph_{guest}$ \\
    \ForEach{$\mathbf{eval}(m_g) \in m_h$ where \\  \label{algo1:1}
      \quad $m_g \in \callgraph_{guest}$ // guest method\\
      \quad $m_h \in \callgraph_{host}$ /* host method */}{
        Extend G to include an edge $(m_h, m_g)$ \\ \label{algo1:2}
        \If{$v \in \mathit{InterfaceVar(m_h)}$ // v is defined in $m_h$ 
        \textbf{and}~$\mathit{v \in use(m_g)}$ // v used in some expression in $m_g$}{ \label{algo1:3}
            $G, B_{m_h} \leftarrow \mathit{DiscoverBridgeCalls}(v, m_h, m_g)$ \\
        }
    }
\end{algorithm}

\cref{algo:callgraph} describes the modular call graph generation. At first we merge the independent host and guest call graphs. The next step identifies guest methods invoked via \emph{eval} in a host method (\cref{algo1:1}), for which an edge is added from the host to the guest method (\cref{algo1:2}). The guest language can also access the host's methods via interface objects. Such flows are added at \cref{algo1:3} (explained in \cref{algo:discover}). The key idea is to search the guest methods for invocations of bridged host methods via interface variables. If such a call exists, edges between the corresponding guest and host methods are added together with edges from the original host method to the bridged host methods. The latter edges are referred to as \emph{bridge call edges}. 
They denote the indirect call relationship to host language methods through bridge objects in the guest language. 
\begin{definition}[Bridge call edge.]\label{def:bridge-call-edg}
	Let $i$ be an interface variable shared from a host method $m_h \in \callgraph_{host}$, $m_g \in \callgraph_{guest}$ such that $m_g$ invokes a bridged method $m_b \in \callgraph_{host}$ via the interface variable $i$ (or its aliases).  Then, a bridge call edge, denoted as $E_{bc}$, is a transitive call edge from $m_h$ to $m_b$ in the interlanguage call graph $G$.
\end{definition}

\Cref{fig:call-graph-example} shows the transitions involved in creating an interlanguage call graph for the example in \cref{lst:toyexample:host}. Step 0 depicts the call graph obtained during the pre-analysis phase. Initially, $\callgraph_{host}$ has no edges originating from \emph{foo} as the pre-analyses is oblivious to edges from foreign function calls. The host call graph is extended to include the bridge methods (step 1). The call graph is expanded further to include edges that perform indirect invocation of guest methods through eval (step 2). In the last step, the call graph is extended to include the bridge call edges (step 3); \emph{setSecret} and \emph{getSecret} are called via the interface variable \emph{i}. Therefore, bridge call edges (\emph{foo}, \emph{setSecret}) and (\emph{foo, getSecret}) are added to $\callgraph$. Bridge call edges are used to determine the methods required for summary-specialization and unification in our analysis.
We define the notion of bridge callgraph as an directed graph formed by bridge edges and nodes in \cref{def:bridge-sub-graph}. 
\begin{definition}[Bridge callgraph\label{def:bridge-sub-graph}]
    A bridge callgraph $B_{m_h}(N', E_{bc})$  is a connected intra-language graph such that $E_{bc}$ comprises all bridge call edges originating in $m_h \in N'$.
\end{definition}

\begin{algorithm}[tb]
	\caption{DiscoverBridgeCalls}\label{algo:discover}
	\KwIn{Interface variable $v$, host language method $m_h$, guest language method $\mathit{m_g}$}
	\KwOut{Interlanguage callgraph $G$}
	\KwOut{Bridge call graph $B_{m_h}$}
		\ForEach{invocation \textit{x.$m()$} in $m_g$ where $x \in \textit{MayAlias}(v)$ \\
		\quad and $m \in M_b$ /* $M_b$ is set of bridge class methods */}{
				add edge $(m_g, m)$ to $G$ \\
				add \textit{bridge call edge} $(m_h, m)$ to $B_{m_h}$ 
		}
	//$\mathit{alias}$ as a parameter to another guest function \\
	\ForEach{invocation $m_g'(*,x,*)$ in $m_g$ where $x \in \textit{MayAlias}(v)$}{
	$\mathit{DiscoverBridgeCalls}(x, m_h, m_g')$
} 
	
\end{algorithm}

The shaded part of \cref{fig:call-graph-example} (step 4) shows the bridge callgraph originating in \emph{foo()} for the example in \cref{lst:toyexample:host} and  \cref{lst:toyexample:guest}. An indirect call from \emph{foo} to \emph{setSecret} and \emph{getSecret} happens through the bridge object $i$. The bridge callgraphs are the core data structures used in summary specialization, unification, and injection. 

\subsection{Modular Analysis Framework}

We now present the analysis's core pillar --  \textit{summary specialization}. Summary specialization resolves the dataflows occurring in the bridge callgraph from the guest to the host and vice versa. The procedure is divided into three steps. Initially, it computes function summaries leveraging the information gained during the pre-analysis. The function summaries are determined using an intra-procedural points-to-analysis. The second step involves merging and propagating these intra-procedural summaries to the functions engaged in the bridge callgraph to resolve the dataflows occurring via the bridge. At this stage, the analysis has identified the dataflows from guest to host through interface variables. These resolved summaries are injected into the corresponding guest methods in the third step. Once we have inserted summaries into the bridge methods, the analysis naturally shifts to the whole program analysis. These techniques are described in detail in \textsc{SpecializeSummary}, \textsc{Unification}, and \textsc{InjectSummary}.

\subsubsection{Function Summary}\label{sec:function-summary-constraints}

\begin{figure}[tb]
$\inferrule[\textbf{Assign}]{x = y}{[y] \subseteq [x]}$ \hfill
$\inferrule[\textbf{Load}]{v = x.f}{[x.f] \subseteq [v]}$\hfill
$\inferrule[\textbf{Store}]{\texttt{x.f = v}}{[v] \subseteq [x.f]}$  \\%\hfill  \phantom{\,}
$\inferrule[\textbf{New}]{\texttt{v = new Obj()} \\ pts(v) = \mathcal{P}(v,\mathit{ctxt})  \\ \mathit{ctxt} = \mathit{CallContext}(v)}{\forall o_i \in pts(v): \{o_i\} \subseteq [v]}$ \hfill 
 $\inferrule[\textbf{Func}]{\texttt{x = func(args)} \\ pts(x) = \mathcal{P}(x,\mathit{ctxt}) \\ \mathit{ctxt} = \mathit{CallContext}(x)}{\forall o_i \in  pts(x): \{o_i\} \subseteq [x]}$  \\
\caption{Constraint system ($\mathcal{C}_s$) for intra-procedural function summaries}
\label{fig:function-summaries-constraints}
\end{figure}

Function summaries in our analysis are obtained through a light-weight intra-procedural pointer analysis. Our function summaries have the form $\mathcal{AP} \mapsto H \cup \emptyset$, where $H$ symbolizes the collection of heap objects, $\mathcal{AP}$ signifies the set of access-paths, and $\emptyset$ indicates that the associated summary has not been resolved. An access path consists of the base variable (quintessentially a local variable) followed by a sequence of field accesses (denoted as \texttt{x.f.g.h}). We choose the access path representation because it inherently models field-sensitivity~\cite{ericbodden-accesspath} and is bounded, making it suited for data-flow analysis~\cite{khedkar-accessgraphs}.

The analysis is based on a straightforward constraint-based approach in which the constraints are reinforced with resolved values from the pre-analysis. We create the function summaries using constraints of an Andersen's style analysis (i.e., inclusion-based)~\cite{SridharanAndersen}. These constraints have been shown to be precise, and they have become the de-facto representation for pointer analysis in programming languages that support heap manipulation~\cite{Sridharan2013}. In general, inclusion-based constraints face scalability issues due to the cubic time and quadratic space complexity on the number of variables; yet, the function summaries in this phase are calculated intra-procedurally and hence confined to a small number of variables.

\begin{algorithm}[tb]
	\caption{\textsc{SpecializeSummary}}\label{algo:compute-function-summary}
	\KwIn{Bridge callgraph $B_{m_h}(N', E_{bc})$}
	\KwIn{Guest pre-analysis $\mathcal{P}_g$ }
	\KwOut{Function summary constraints $\mathcal{F}$}
	\ForEach{function node $N_i \in N'$ \label{algo:compute-function-summary:3}}
	{
        \Comment{Generate constraints from constraint system in \cref{fig:function-summaries-constraints}} \\
		$\mathcal{C}^N_i \leftarrow \mathit{MakeConstraints(N_i, \mathcal{C}_s)}$ \label{algo:compute-function-summary:4}  \\
        \ForEach{guest function node $N_g$ invoking $N_i(\bar{q})$ (calling bridge method) /* $\bar{q}$ is list of formal parameters */}{ \label{algo:compute-function-summary:5}
            let $N_i(\bar{t})$ be the function invocation in $N_g$ // $\bar{t}$ is the list of actual parameters \\
            add $\mathcal{P}_{g}(t_j) \subseteq [q_j]$ to $\mathcal{C}^N_i$ where $t_j \in \bar{t}, q_j \in \bar{q}$ \label{algo:compute-function-summary:6}\\
        }
        \ForEach{$c_i \in \mathcal{C}^N_i$}{
            \If{$c_i$ contains \texttt{this.fld}}{ 
                replace \textit{this} in $c_i$ by the interface variable $b$  
            }
        } \label{algo:compute-function-summary:7}
	
	$\mathcal{F}^N_i = \mathit{\mathsc{CubicSolver}(\mathcal{C}^N_i)}$  \label{algo:compute-function-summary:11} \\
	$\mathcal{F} = \mathcal{F} \cup \mathcal{F}^N_i $ \label{algo:compute-function-summary:12}
} \label{algo:compute-function-summary:13}
$\textsc{UnifySummary}(\mathcal{F}, B_{m_h})$ \label{algo:compute-function-summary:14}
\end{algorithm}

\begin{figure}[tb]
    \centering
    \includegraphics[width=\columnwidth]{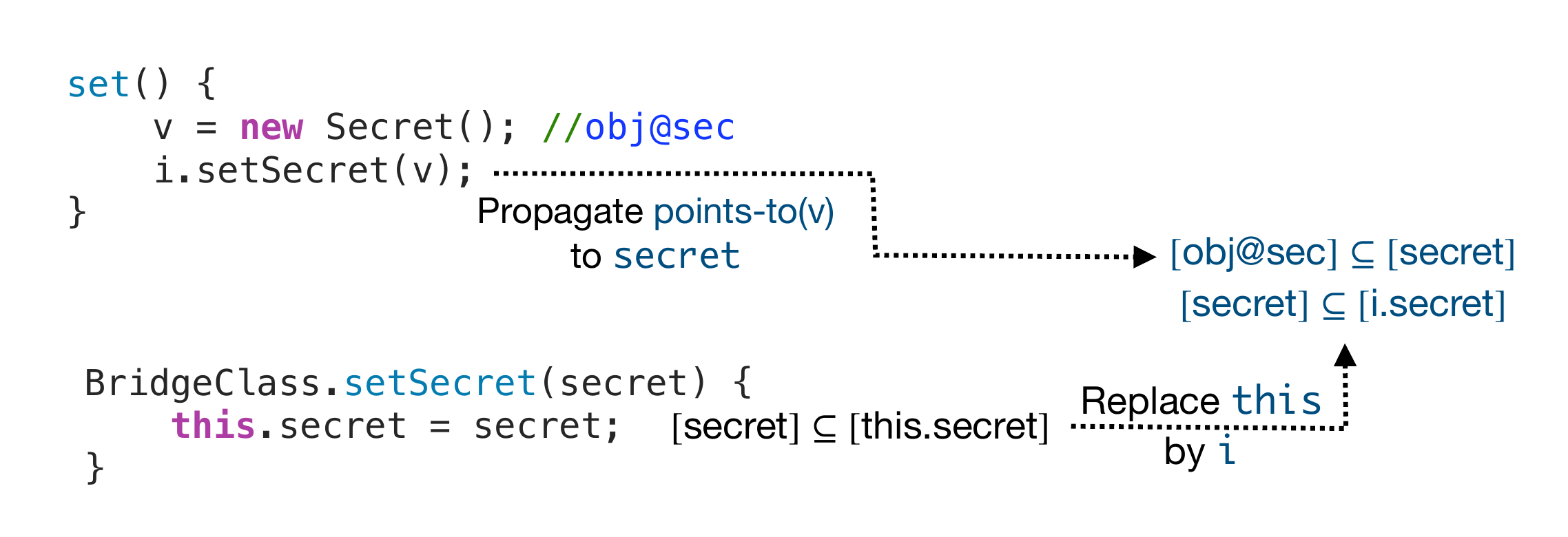}
    \caption{Function summary constraints for \texttt{setSecret}}
    \label{fig:summary-constraint-setsecret}
\end{figure}

\cref{fig:function-summaries-constraints} shows the constraint system to generate a function summary. We use the symbol $[var]$ to denote the constraint variable for $var$, $\mathcal{P}$ denotes the pre-analysis, and $pts(var)$ denotes the points-to set for variable $var$. \textsc{Assign},  \textsc{Store}, and \textsc{Load} are standard Andersen style constraints. The rule \textsc{New} queries the points-to set of the variable from the pre-analysis and constructs a constraint for each heap object. In the case of function calls \textsc{Func} directly obtains the points-to set of the result-variable (capturing the return value) from the pre-analysis, which already contains inter-procedural analysis. The collected constraints are then resolved to function summaries.  \cref{algo:compute-function-summary} illustrates the procedure to acquire function summary constraints. Given a bridge callgraph and the pre-analysis, it first generates the function summary constraints for the bridge callgraph's function nodes. Note that (except for the root) the function nodes in a bridge callgraph comprise the (bridged) functions of the host, which provide functionality to the guest. After obtaining the constraints for these functions, the next step is to map their formal parameters to the actual invocations' parameters in the guest. \cref{fig:summary-constraint-setsecret} illustrates the procedure from \cref{algo:compute-function-summary:5} to \cref{algo:compute-function-summary:7}. These constraints are then sent into a cubic solver, which generates the actual function summary constraints.

\begin{figure}[tb]
    \small
    \begin{tabular}{lll}
        \toprule
        \textbf{Function} & \textbf{Constraints} & \textbf{Function Summary} \\
        \midrule
        \textsf{foo} & $\{\mathit{obj@iface}\} \in [\mathit{i}]$ & $[\mathit{i}] \mapsto \{\mathit{obj@iface}\}$ \\
                       & $ \emptyset \subseteq [x]$ & $[\mathit{x}] \mapsto \emptyset$ \\
        \midrule
        \textsf{setSecret} & $[\mathit{secret}] \subseteq [\mathit{i.secret}]$ & $[\mathit{i.secret}] \mapsto \{\mathit{obj@sec}\}$   \\
                            & $\{\mathit{obj@sec}\} \in [\mathit{secret}]$ & \\
        \midrule
        \textsf{getSecret} & $[\mathit{i.secret}] \subseteq [\mathit{\$ret}]$ & $[\mathit{i.secret}] \mapsto \emptyset $ \\
            & &  $[\mathit{\$ret}] \mapsto \emptyset$ \\
        \bottomrule
    \end{tabular}
\caption{Summary constraints and resolved summaries}
\label{fig:function-summaries-constraints-resolved}
\end{figure}

\cref{fig:function-summaries-constraints-resolved} shows the constraints and function summaries generated for our example program. The second column shows the constraints and the third column  the resolved summaries after the cubic solver. In the function \textsf{foo}, we have a single constraint $\{\mathit{obj@iface}\} \subseteq [\mathit{i}]$ which is resolved to $[\mathit{i}] \mapsto \{\mathit{obj@iface}\}$. In the function \textsf{setValue}, we have two constraints $\{\mathit{obj@sec}\} \in [\mathit{secret}]$ and $[\mathit{secret}] \subseteq [\mathit{i.secret}]$;  they are resolved to $[\mathit{i.secret}] \mapsto \{\mathit{obj@sec}\}$. In the function \textsf{getValue}, we have a single constraint $[\mathit{i.secret}] \subseteq [\mathit{\$ret}]$ resolved to $[\mathit{\$ret}] \mapsto \emptyset$ and  $[\mathit{i.secret}] \mapsto \emptyset$.

\subsubsection{Unification}\label{sec:summary-unification}

As evident in \cref{fig:function-summaries-constraints-resolved}, in the \textsf{setSecret} function \textsf{i.secret} corresponds to the guest object \emph{obj@sec}; however, in the \textsf{getSecret} function, \textsf{i.secret} remains unknown, despite the fact that the access paths are identical based on the same interface variable $i$. To address this, we combine these summaries according to the access path. 

\begin{figure}[tb]
    \centering
    \includegraphics[width=\columnwidth]{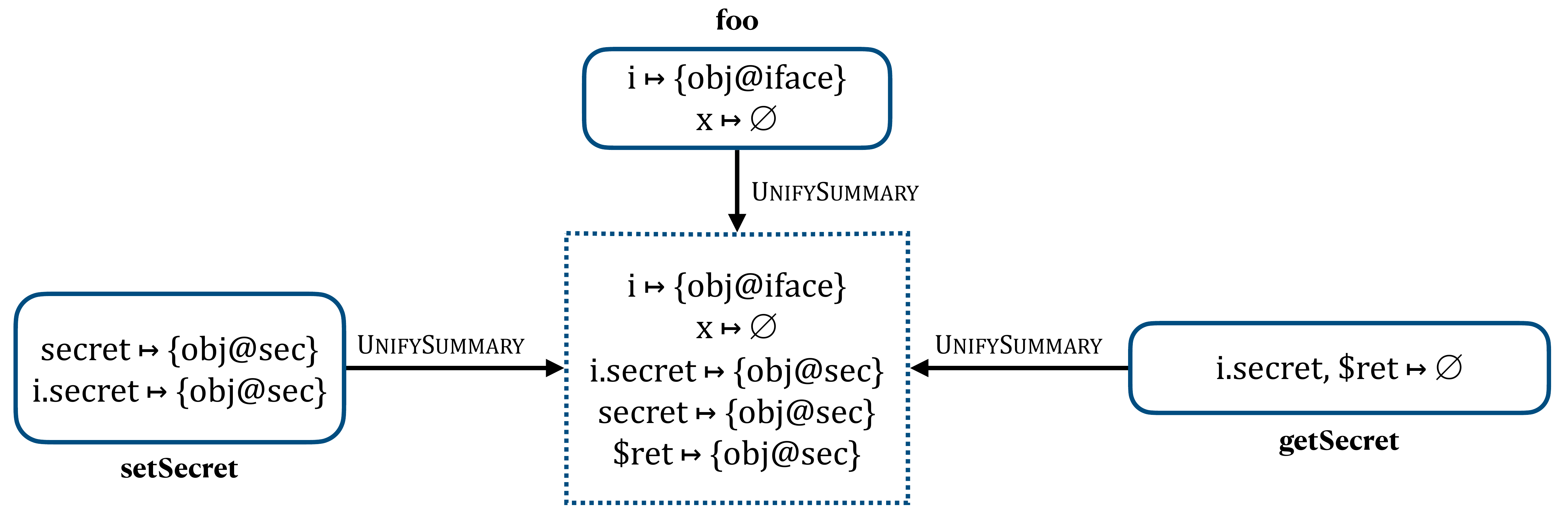}
    \caption{Unification for the bridge callgraph in \cref{fig:call-graph-example}}
    \label{fig:unification-bridge-subgraph}
\end{figure}

Two summaries with the same access path are combined by a union operation, i.e., if $ap \mapsto \phi_1$ and $ap \mapsto \phi_2$, then the unification results in $ap \mapsto \phi_1 \cup \phi_2$. The unification is also alias-aware, i.e., aliases of the base variable in $ap$ are also unified. For simplicity, we assume that the access paths are not nested. This is a fair assumption since most dataflow analyses are performed on an intermediate language, which simplifies the language to such a syntax for analysis. \cref{fig:unification-bridge-subgraph} shows the result of unification for the bridge callgraph shown in \cref{fig:call-graph-example}. This step unifies the summaries based on the access paths and aliased access paths. First, the analysis resolves the data flow across bridge functions by comparing the access paths, and then it updates the corresponding points-to sets. Therefore, $\mathit{i.secret}$ points to the guest object $\mathit{obj@sec}$ across all functions participating in the bridge callgraph.
        
\subsubsection{Summary Injection}\label{sec:summary-injection}

After summary unification, the analysis has resolved the data flows across the bridge callgraph. The analysis, in particular, resolved the operations in bridge methods, i.e., host language functions called from the guest interface. Two tasks remain: First, translating the analysis summaries to be compatible with the analysis models adopted by the host and guest languages. Second, propagating the translated summaries to the invoking functions at the guest side.

 The summaries obtained via unification are in the access path format; however, we need to translate them to the host and guest languages' analysis models to reuse them in the host and guest. Notably, we translate the access path format to their native format. The two levels of sensitivities involved in the underlying analysis, calling context and field sensitivity, are of particular relevance. Since the bridge functions are invoked through interface variables in the guest applications, the context of the host method is assigned in the bridge callgraph. Following that, we classify field sensitivity according to how the underlying analysis handles field operations; as field-sensitive, field-based, or field-insensitive. The analysis determines the appropriate field sensitivity representation depending on the type of underlying analysis.

Considering the running example, the analysis resolves that the \textsf{getSecret} function returns $\texttt{obj@sec}$. The return value is captured by the variable $g$ in \textsf{getAndLeak}; however, at this stage $g$ has not been resolved and still points to $\emptyset$. Consequently, the analysis misses that $g$ exposes a sensitive value at \cref{lstlistingA:15}. Summary injection performs this step by retrofitting the pre-analysis with the unified summary. First, the unified summary information is translated to the host language’s analysis model. The pre-analysis updates cause an incremental update of the variables impacted by the injected dataflows. Next, these updates (translated summaries) are propagated to the affected guest functions.

\begin{algorithm}[tb]
    \caption{\textsc{InjectSummary}}
    \label{algo:inject-summary}
    \KwIn{Projection $\mathcal{R}^f$,  pre-analysis $\mathcal{P}$ }
    \KwIn{Function Summary $\mathcal{F}^f$ of a function $f$ }
    \KwOut{updated  pre-Analysis $\mathcal{P}_{modular}$}
    $\mathit{changedVars} \leftarrow \emptyset$ \\
   Updated Projection $\mathcal{R}^f_\Delta\leftarrow \emptyset$ \\
    \ForAll{$(\mathit{accesspath}, \mathit{heap}) \in \mathcal{F}^f$}{ \label{algo:inject-summary-loop-begin}
        \If{$\mathit{accesspath}$ is a simple variable $\mathit{var}$}{
            $\mathcal{R}^f_\Delta = \mathcal{R}^f_\Delta \cup \{(\mathit{var}, \mathit{heap}, \mathit{ctxt})\}$ for any $\mathit{ctxt} \in \mathcal{R}^f$  \label{algo:inject-summary-simple-var}  \\
        }
        \If{$\mathit{accesspath}$ is a field access x.f}{
            
                \Case{\textsc{FieldSensitive} analysis} {
                    $\mathcal{R}^f_\Delta = \mathcal{R}^f_\Delta \cup \{(o_i,f,heap) \mid \forall o_i \in  \mathcal{P}(x)$\}  \label{algo:inject-summary-field-sensitive} \\
                }
                \Case{\textsc{FieldBased} analysis}{
                    $\mathcal{R}^f_\Delta = \mathcal{R}^f_\Delta \cup \{(f,heap)\}$ \label{algo:inject-summary-field-based} \\
                }
                \Case{\textsc{FieldInsensitive} analysis}{
                    $\mathcal{R}^f_\Delta = \mathcal{R}^f_\Delta \cup \{(o_i, heap) \mid \forall o_i \in \mathcal{P}(x)\}$ \label{algo:inject-summary-field-insensitive} \\
                }

        }
        $\mathit{changedVars} \leftarrow \mathit{changedVars} \cup \{\mathit{accesspath}\}$ \\
    } \label{algo:inject-summary-loop-end}
// Update the guest projection to include resolved host projection \\
    \ForEach{$f' \in G_{\mathit{guest}}$ invoking $f$, which is a bridge method} 
    {
    	// Propagate host function's ($f$) return variable to guest ($f'$), adapting variable name. \\
    $	\mathcal{R}^{f'} = 	\mathcal{R}^{f'} \cup \mathcal{R}^f_\Delta[\$ \mathit{ret}]$ \\
        
	}

  $\mathcal{P}_{modular} = \mathcal{P} \cup \mathcal{R}^{f'}$
\end{algorithm}
\cref{algo:inject-summary} shows the steps for summary injection. For a given function $f$, it takes the analysis snapshot of $f$ as an input. We denote this snapshot as a \emph{projection} of a function. Formally, we define the \emph{projection} as:
\begin{definition}[Projection]\label{def:projection}
	A projection $\mathcal{R}^f = \bigcup%
	_{v \in \mathit{def}(f)}\{(v, h, \mathit{ctxt}) \mid h = \mathit{pointsTo} (v, \mathit{ctxt})\}$ for a function $f$ is a set of points-to sets for the variables defined in the function $f$ under a context $\mathit{ctxt}$.   
\end{definition}
For the function $f$ in the bridge graph, our analysis gets the function's context and produces a projection of the analysis. Intuitively, projection is the points-to sets of the variables defined in the function for a context $\emph{ctxt}$. \cref{algo:inject-summary} uses the projection of the analysis to inject the dataflow facts obtained from unification. Along with the projection, the algorithm accepts the function summary of a function $f$ in access-path format, and information on how the function  handles field operations --- either \textsc{FieldSensitive} or \textsc{FieldBased} or \textsc{FieldInsensitive}. Earlier research~\cite{SridharanAndersen, sridharan13efficient, ericbodden-pds,Yannis2017-OOPSLA-PTaint, YannisOOPSLA2009, object-sensitive-milanova} shows that these are the primary techniques for handling field operations, and hence, we confine our algorithm to those.
\cref{algo:inject-summary-loop-begin} to \ref{algo:inject-summary-loop-end} translate the obtained summary information in access-path representation to the analysis-supported representation. In \cref{algo:inject-summary-simple-var}, the translated points-to-set for the simple variables are added to the projection. \cref{algo:inject-summary-field-sensitive} incorporates the translated points-to-set for field-sensitive analysis, \cref{algo:inject-summary-field-based} for the field-based variables, and  \cref{algo:inject-summary-field-insensitive} for the field-insensitive variables. In the following step, the modified host projection (of the bridge method) is propagated to the calling guest method (connected via bridge method edge in bridge graph). Finally, the pre-analyses are revised and translated projections are included. With these revisions, the analysis has successfully resolved all key operations for (one) bridge callgraph's inter-language operation.

\begin{figure}
	\centering
	\begin{subfigure}[b]{\linewidth}
		\includegraphics[width=1.05\textwidth]{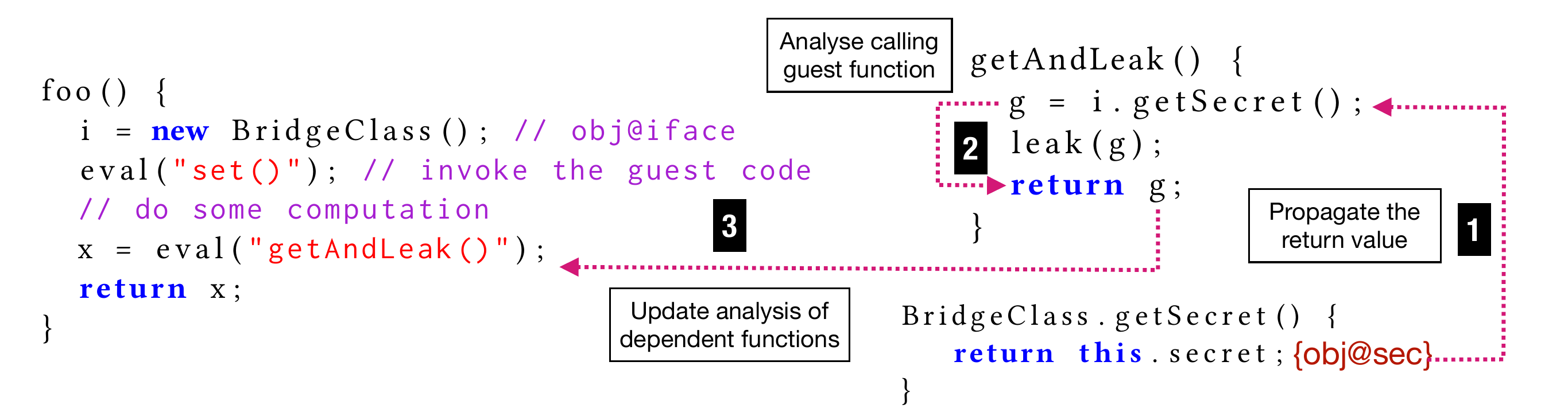}
		\caption{Summary Injection in methods}
		\label{fig:analysis-visualization}
	\end{subfigure}
	\begin{subfigure}[b]{\linewidth}
		\centering
		\begin{adjustbox}{width=\columnwidth,center}
\begin{tabular}{ccc}
    \toprule
    \textbf{Method} & \textbf{Before Summary Injection} & \textbf{After Summary Injection} \\ \midrule
    \textsf{foo} & $\langle i, ctx \rangle \mapsto \{\texttt{obj@iface}\}$ & $\langle i, ctx \rangle \mapsto \{\texttt{obj@iface}\}$ \\
                 & $\langle x, ctx \rangle \mapsto \emptyset$ & \textcolor{red}{$\langle x, ctx \rangle \mapsto \{\texttt{obj@sec}\}$} \\ \midrule
    \textsf{setSecret} & $\langle i.secret, ctx' \rangle \mapsto \{\texttt{obj@sec}\}$ &  $\langle i, secret, ctx \rangle \mapsto \{\texttt{obj@sec}\}$ \\ \midrule
    \textsf{getSecret} & $\langle i.secret, ctx' \rangle \mapsto \{\texttt{obj@sec}\}$ &  $\langle i, secret, ctx \rangle \mapsto \{\texttt{obj@sec}\}$ \\ 
                       &  $\langle \mathdollar ret, ctx' \rangle \mapsto \{\texttt{obj@sec}\}$ &  $\langle \mathdollar ret, ctx \rangle \mapsto \{\texttt{obj@sec}\}$ \\ \midrule
    \textsf{set} & $\langle v, ctx' \rangle \mapsto \{\texttt{obj@sec}\}$ & $\langle i.secret, ctx' \rangle \mapsto \{\texttt{obj@sec}\}$ \\ \midrule
    \textsf{getAndLeak} & $\centering \emptyset$ &  \textcolor{red}{$\langle g, ctx' \rangle \mapsto \{\texttt{obj@sec}\}$} \\ \bottomrule

\end{tabular}
\end{adjustbox}
		\caption{Analysis before and after summary injection}
	\end{subfigure}
	\caption{Summary Injection for \cref{lst:toyexample:host} and \cref{lst:toyexample:guest}}
		\label{fig:summary-injection-example}
\end{figure}

\cref{fig:summary-injection-example} summarizes the output of the summary injection to the pre-analysis. The second and third columns depict a snapshot of the analysis prior to and following the summary injection. The return value of the getSecret method has been updated to $\{\texttt{obj@sec}\}$ as a result of the summary unification. Since \textsl{getAndLeak} invokes the function \textsl{getSecret}, the summary injection automatically propagates the updated return value to \textsl{getAndLeak}, and the points-to set of \textsl{g} now contains $\{\texttt{obj@sec}\}$. The bridge callgraph includes the connection from \textsl{getAndLeak} to \textsl{foo}, and consequently, the $\{\texttt{obj@sec}\}$ is propagated to the variable \textsl{x}.

\subsubsection{Overall Workflow}\label{sec:putting-all-together}
\Cref{algo:stich-analysis} shows the workflow of the algorithms mentioned before. Our algorithm runs by computing the call graphs and unifying the pre-analysis via summary specialization until it no longer changes, i.e., achieves a fixed point.

\begin{algorithm}[tb]
\caption{\mathsc{PointerAnalysisSummarySpecialization}}
\label{algo:stich-analysis}
\KwIn{$\mathcal{P}$ Pre-analyses}
\KwIn{$G_{\mathit{host}}, G_{\mathit{guest}}$ host and guest language callgraphs}
\KwOut{Analysis $\mathcal{P}_\mathit{modular}$}
    
\Repeat{$\mathcal{P}_\mathit{modular}$ is unchanged (i.e., reaches a fixed point)}{
    $G, B \leftarrow \textsc{ModularCallgraphAnalysis}(G_{\mathit{host}}, G_{\mathit{guest}} )$ \\
    $\mathcal{F} \leftarrow \textsc{SpecializeSummary}(B, \mathcal{P}_g)$ \\
    \ForAll{$f \in \mathit{functions(B)}$}{
        $\mathcal{R}^{f} \leftarrow \mathit{Projection}(f)$ \\
        $\mathcal{P}_\mathit{modular} \leftarrow \mathsc{InjectSummary}(\mathcal{R}^{f}, \mathcal{P},  \mathcal{F}^f)$ \label{algo:analyzebridgecalls:injection}\\
    }
}
\end{algorithm}

\subsubsection{Multiple Bridge Calls}\label{sec:multiple-bridge-calls}
\begin{figure}
	\centering
	\includegraphics[scale=0.5]{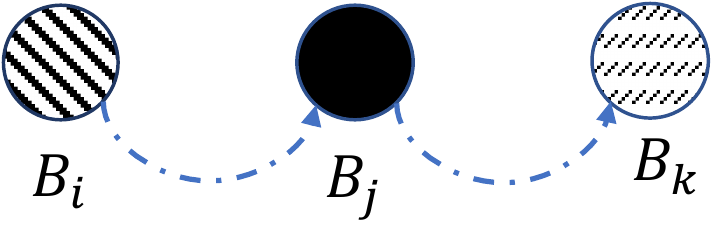}
	\caption{Multiple Bridge callgraphs}
	\label{fig:multiple-bridge-subgraphs}
\end{figure}
As is, bridge calls originating from different host functions will produce multiple bridge callgraphs, thus, (potentially) generating a list of disconnected bridge callgraphs, as illustrated in \cref{fig:multiple-bridge-subgraphs}. To aid in the pre-analysis of actual data flow across bridge callgraphs, resolved summaries across each bridge callgraph should propagate in the correct sequence to others. In our approach, we use the traditional topological ordering, i.e., the execution order as generated by the pre-analysis. In \cref{fig:multiple-bridge-subgraphs}, for example, analysis resolves the data flow for the bridge callgraph $B_i$, and then, summary injection propagates the modifications down the call-graph (to $B_j$ and subsequently to $B_k$).

\subsubsection{A Typical Dataflow Analysis}\label{sec:data-flow} 
Given a traditional security lattice, the core part of our framework works with the points-to set where variables are annotated with the security labels. Once the annotated points-to-sets have been resolved, a taint analysis can query those for the alias sets and subsequently use them for taint propagation~\cite{boomerang, hybridprunning, Yannis2017-OOPSLA-PTaint}. 
A detailed information flow/taint analysis is accompanied as supplementary material.

\section{Limitations}

\Cref{fig:language-model} only allows programs as parameters to eval, so no dynamically computed target. In practice this is not a restriction in contemporary programs but could be lifted with additional analysis~\cite{grech_et_al:LIPIcs:2018:9231,10.1145/3426470}. 
Our approach uses may aliases of the interface object on the guest side. Thus, our analysis may yield imprecise results when a bridge method has the same name and arity as the guest method. However, this is a general limitation of static analysis; besides, we have not seen such occurrences in our dataset. 
The language assumes that \emph{bridge interfaces} are specified and their type is known to the analysis to identify \emph{InterfaceVars}. %
This is a fair assumption as modern languages specify the syntax and semantics of these markers, e.g., Android defines WebView APIs for bridge interface. Considering this, our analysis assumes that methods invoked via interface objects are statically resolved, i.e., dynamic binding is not applicable to the bridge class’s methods invoked via interface objects. Our analysis obtains these statically resolved interface objects from the pre-analysis.

\section{Implementation}
We developed a prototype implementation of our approach called~\toolname. \toolname~currently supports analysis of Java-C programs and Java-Python programs. \toolname~is implemented in Java, consists of about 5k lines of code, and builds on top of the WALA analysis framework~\cite{wala} for analyzing Java and Python. For analyzing C/C++ programs, we choose SVF~\cite{svf, sui2014detecting}, a popular tool for C/C++ dataflow analysis. 

\textbf{Extensibility:} At the time of writing this paper,~\toolname~ supports WALA-SVF and Java-Python analysis in Wala. However, \toolname~is extensible to other language combinations (with some implementation efforts for language bindings). \toolname~provides a set of interfaces (Adaptors) to extend to other analysis frameworks. A detailed API document containing the Interface details will be open-sourced along with  \toolname.
 
For analyzing Android NDK programs, \toolname~takes the android apk file as input. It uses the \emph{retdec} decompiler~\cite{avast} to lift the NDK libraries bundled in the apk file to LLVM. It takes the Java jar files and Python sources as input for Graal Polyglot APIs.

\section{Evaluation}

We empirically evaluate~\toolname’s efficacy in resolving the inter-language operations in polyglot applications. To demonstrate the generalizability of~\toolname~to other inter-operable languages, we evaluate~\toolname~on two polyglot language models: (1) Java-C communication in Android NDK apps and (2) Java-Python applications using GraalVM. Java-C communication is manifested in Java (or Android) applications that use system-level features for various reasons, such as performance or interfacing with the hardware. Java-Python applications in our dataset use the GraalVM Polyglot API. In particular, our evaluation is designed to address the following research questions:
\begin{enumerate}[leftmargin=*,labelindent=\parindent]
	\item[\textbf{RQ1}] How does~\toolname~compare to the state-of-the-art polyglot analyses in terms of precision and scalability?
	\item[\textbf{RQ2}] Can~\toolname~generalize to multiple polyglot language models?
\end{enumerate}
All the experiments were performed on a %
laptop running macOS Monterey with a 10-Core M1 Pro processor and 32~GB RAM.

\subsection{RQ1: Comparison to the state-of-the-art}
Recent years have seen a few research tools~\cite{Ryu-Semantic-Summary,Monat-Python-SAS,weiJNSAF-CCS2018,buro2020abstractmultlingual} to address the analysis of polyglot apps. C-summary~\cite{Ryu-Semantic-Summary} is the most recent multilingual analysis tool, and it is shown to overcome various limitations of previous works. Thus, we compare the effectiveness of~\toolname~with C-summary~\cite{c-summary-github}.

\textit{Subject apps.} To evaluate the first research question, our criteria for selecting the subject apps is dictated by two aims: (1) include apps from the C-summary dataset to avoid selection bias, and (2) evaluate~\toolname~on various Java and C communication patterns. Considering these two choices, we compiled 23 Android JNI apps from the widely used NativeFlowBench~\cite{NativeFlowBench, weiJNSAF-CCS2018} and 16 popular apps from the Google play store. NativeFlowBench is an Android app benchmark that includes various inter-language communication features. Each app in this benchmark has a ground truth.

 Next, we assess all tools on the following parameters: (1) whether it can discover the callgraph edges from the host-to-guest language and vice-versa, and (2) whether it can discover the reachable methods in the guest language to generate a whole-program call-graph.

\newcommand{\xmark}{\ding{55}}%
\newcommand{\cmark}{\ding{51}}%

\begin{table*}
    \centering
    \caption{Analysis of JNI from benchmark from \textsf{NativeFlowBench} microbenchmark~\cite{NativeFlowBench} and Real-world JNI Apps. $\textsf{T}$ means that the the underlying tool synthesizes an imprecise program such as \texttt{Java.Top()}. $\square$ denotes that the tool gives empty programs.}
    \label{table:JavaCEvaluation}
\begin{adjustbox}{width=0.7\linewidth}
\begin{tabular}{|l|c|ccccccc|}
\hline 
\textbf{Benchmark} & \textbf{C-Summary} & \multicolumn{7}{c|}{\textbf{Conflux}} \\ 
                   & \textbf{Generated} &   \textbf{Analyze} & $N_J$ & $E_J$ & $N_C$ & $E_C$ & IL & Time (in sec) \\ \hline \hline
\multicolumn{9}{|c|}{\textbf{NativeFlowBench}} \\ \hline
complexdata & $\square$ & \cmark & 364 & 940 & 8 & 10 & 2 & $< 1$ \\ \hline
complexdata stringop & \xmark & \cmark & 5802 & 13015 & 9 & 10 & 1 &  $< 1$ \\ \hline
dynamic register multiple & \xmark & \cmark & 382 & 979 & 9 & 13 & 3 & $< 1$ \\ \hline
heap modify & $\square$ & \cmark & 6889 & 15272 & 11 & 12 & 1 &  $< 1$ \\ \hline
icc javatonative & \xmark  & \xmark & 366 & 895 & 13 & 14 & 0 & $< 1$ \\ \hline
icc nativetojava & $\square$ & \cmark & 7865 & 14891 & 8 & 9 & 1 &  $< 1$ \\ \hline
leak & $\square$ & \cmark & 367 & 894 & 3 & 4 & 1 & $< 1$\\ \hline
leak array & $\square$ & \cmark & 383 & 968 & 4 & 5 & 1 & $< 1$\\ \hline
leak dynamic register & $\square$ & \cmark & 367 & 895 & 9 & 11 & 1 & $< 1$ \\ \hline
method overloading & $\square$ & \cmark & 307 & 751 & 3 & 5 & 2 & $< 1$ \\ \hline
multiple interactions & $\square$ & \cmark & 384 & 980 & 11 & 13 & 2  & $< 1$\\ \hline
multiple libraries & $\square$ & \cmark & 382 & 978 & 3 & 5 & 1 & $< 1$\\ \hline
noleak & $\square$ & \cmark & 367 & 894 & 1 & 2 & 1  & $< 1$\\ \hline
noleak array & $\square$ & \cmark & 5795 & 13008 & 4 & 5 & 1 & $< 1$ \\ \hline
nosource & $\top$ & \cmark & 306 & 746 & 1 & 2 & 1  &  $< 1$\\ \hline
pure & \xmark & \xmark & 6581 & 14686 & 39 & 42 & 0  &  $< 1$\\ \hline
pure direct & \xmark & \xmark & 6581 & 14686 & 30 & 48 & 0  & $< 1$\\ \hline
pure direct customized & \xmark & \xmark & 380 & 958 & 30 & 48 & 0 &  $< 1$ \\ \hline
set field from arg & $\top$ & \cmark & 373 & 908 & 4 & 5 & 1 & $< 1$ \\ \hline
set field from arg field & $\top$ &  \cmark & 314 & 763 & 4 & 5 & 1  & $< 1$\\ \hline
set field from native & $\top$ & \cmark & 310 & 762 & 14 & 15 & 1 &  $< 1$ \\ \hline
source & $\top$ & \cmark & 6892 & 15271 & 7 & 8 & 1 & $< 1$ \\ \hline
source clean & $\top$ & \cmark & 385 & 985 & 4 & 5 & 1 &  $< 1$ \\ \hline \hline

\multicolumn{9}{|c|}{\textbf{Google PlayStore Apps}} \\ \hline
Airmessage & - & \cmark & 53987 & 30886 & 169 & 141 & 4 & 30  \\ \hline
Apple Music & -  & \cmark & 51867 & 28760 & 8334 & 3055 & 8 & 72 \\ \hline
Arte & - & \xmark & 57224 & 31237 & 74 & 31 & 0 & 480  \\ \hline
ClearScore & - & \xmark & 50834 & 31063 & 344 & 321 & 0 & 30 \\ \hline
CommonsLab & - & \cmark & 28817 & 16576 & 447 & 348 & 14 & 15  \\ \hline
FPLayAndroid &  $\top$ & \cmark & 5052 & 2316 & 294 & 238 & 14 & 240 \\ \hline
FairEMail &  $\top$ & \cmark & 59766 & 40214 & 3854 & 1116 & 5 & 85 \\ \hline
JioMoney & - & \cmark & 50002 & 22904 & 299 & 232 & 2 & 27 \\ \hline
NetGuard & - & \cmark & 12159 & 6448 & 333 & 168 & 2 & 5 \\ \hline
PCAPdroid & $\top$ &  \cmark & 54270 & 28835 & 613 & 303 & 3 & 29 \\ \hline
Proton VPN & $\top$ & \cmark & 56217 & 31253 & 56217 & 8 & 1 & 54 \\ \hline
Spotify & - & \xmark & 59690 & 33736 & 439 & 177 & 0 & 600  \\ \hline
Spotify Lite & -  & \cmark & 51728 & 31234 & 843 & 434 & 8 &  390 \\ \hline
TermOne & $\top$ & \xmark & 47352 &  24365 & 48 & 30 & 0 & 22 \\ \hline
Termux & $\top$ &\cmark & 15656 & 9603 & 77 & 36 & 6 & 6 \\ \hline
TimidityAE & $\top$ & \cmark & 15278 & 8275 & 187 & 127 & 6 & 1 \\ \hline

\end{tabular}
\end{adjustbox}
\vspace{-1em}
\end{table*}

\Cref{table:JavaCEvaluation} contains the subject apps and shows the evaluation results of~\toolname~and C-Summary on them. The first column shows the subject apps' name, and the second includes the analysis results obtained from C-summary. The columns $N_C$ and $E_C$ denote the number of nodes and edges in call graph of the native component, column $\mathit{IL}$ for each tool lists the number of edges in the call graph that go from the host language to the guest language. \toolname~was configured for one-callsite-sensitive and field-sensitive analysis for Java and field-sensitive context-insensitive analysis for C/C++.

\paragraph*{Our Experience with C-summary}
\Cref{table:JavaCEvaluation} uses a few symbols to denote the evaluation results of~\toolname~and C-summary. The symbol $\square$~indicates that the analysis was able to detect the function signature but failed to synthesize the function definition. In our experiments, in 11 out of 19 apps, C-summary failed to synthesize a function definition. This happens when a program has complex pointer operations (such as pointer arithmetics) or method overloading; for example, programs use string library operations such as \texttt{strcpy}, which do not have the source code. The symbol $\top$~indicates the tool could identify the function signature but generated an empty program such as \texttt{Java.Top()}, which essentially returns the set of all heap allocations (imprecise analysis result). In six out of 19 apps, C-summary generated an empty program. This happens when object field load and store methods are accessed in native code. For example, in the benchmark app \emph{complexdata\_stringop}, C-summary generates an empty program for the Java methods \texttt{GetFieldID} and \texttt{GetObjectClass}. The symbol \xmark~indicates that a tool could neither identify the function signature nor synthesize the program. In six out of 19 apps, C-summary showed this behavior. This happens in two cases: (1) intent operations over native classes and (2) operations not handled by C-summary such as arrays or global variables, which are reported limitations of their framework~\cite{Ryu-Semantic-Summary}. Thus, C-summary could not synthesize the required program for further analysis by Android analysis tools, thus hiding the functions reachable through the bridge interfaces. In many cases it could only identify the function definitions used to find reachable native functions from the Java (Android) side.

\paragraph*{\toolname's Results}
\toolname~uses SVF (for C/C++) and Wala (for Java) as the pre-analyses. We configured SVF to run with the entrypoints defined in the host language analysis. Based on the entrypoints, it triggers SVF to analyze the guest functions in the environment and collects the results. Consequently, SVF identifies the reachable functions in the guest language and \toolname~combines the analysis result with those obtained from Wala to build an inter-language call graph. The columns $N_{J}$ and $E_{J}$ show the number of call graph nodes and edges identified by the respective pre-analyses. The numbers are significantly high considering the size of the benchmarks because Wala includes a number of libraries by default~\cite{PointEval}, which increases the call graph size. The column \textsf{IL} reports the interlanguage edges in the call graph, i.e., the edges which go from Java to C and vice-versa. Column $E_{C}$ shows the edges identified by SVF. In our evaluation, \toolname~is able to identify the inter-language edges in 19 out of 23 benchmarks, i.e., it identifies the methods which would have been previously reported unreachable with Wala.
In the remaining four cases, \toolname~does not identify the entry point where native code is invoked via intents, which is expected as middleware-based inter-language communication is out of the scope of this study. Intent resolution tools~\cite{IccTA, Amandroid, iifa} could supplement \toolname~to resolve these entrypoints.

\begin{mdframed}[backgroundcolor=gray!20, roundcorner=20pt]
\textbf{RQ1(a):} In 19 out of 23 subjects, \toolname~builds the complete inter-language call graph. \toolname~leverages the strength of individual analyses and maps required call edges from host-to-guest and vice-versa.
\end{mdframed}

\paragraph*{Scalability} To evaluate the efficacy of~\toolname~on real-world apps, we downloaded 50 popular and recent apps from the Google play store. To filter out NDK apps, we vetted their decompiled code for the presence of shared object files, resulting in 16 apps. These apps include popular apps such as Spotify, and Apple Music. We share a few apps in our dataset with those evaluated by C-summary.

The first part of the bottom half of \Cref{table:JavaCEvaluation} shows the results obtained on these apps from C-Summary. Since C-summary is designed to run on source code, we ran it on seven apps that are open-source. In all of these apps, C-summary synthesized imprecise programs (statement containing \texttt{Java.Top()}).

The second part of \Cref{table:JavaCEvaluation} shows the results of our analysis on these apps. Our analysis scales for many popular apps and analyzes relatively large apps such as \emph{Spotify} within 10 minutes (median around 2 minutes). Among these,~\toolname~was able to identify inter-language call graph edges in 12 out of 16 apps. Again, in the other four apps, guest functions were called through intents, which are out of scope. To validate the number of inter-language edges, we scanned the decompiled files for the presence of native markers in the binary. Two authors independently validated these results.

\begin{mdframed}[backgroundcolor=gray!20, roundcorner=20pt]
    \textbf{RQ1(b):} In 12 out of 16 large apps, \toolname~successfully builds the interlanguage call graph in  about 2 minutes on average. It demonstrates that combining the individual analyses via~\toolname~is a promising direction for a scalable dataflow analysis in polyglot applications. 
\end{mdframed}
\subsection{RQ2: Analysis of Java-Python programs with GraalVM}
To evaluate the generalizability of our approach, we applied \toolname~to Java-Python-based GraalVM programs. GraalVM's Polyglot API is relatively new, and we could not find many projects using Java-Python-based communication on GitHub. To curate the apps, we selected projects with the keyword \textsf{graalpython} and those importing the class (or subclasses) of \textsf{org.graalvm.polyglot.Context}. Out of the seven results we obtained on Github, we searched for the presence of Python code snippets and files. Finally, we confirmed two applications with Java as the host language and Python as the guest (\cref{java-pyhton}). We also develop an additional microbenchmark for analysis, in total, having three micro-benchmarks for study. In BM-1~\cite{bm1}, Java setter and getter methods are invoked from  Python. BM-2~\cite{bm2} invokes Java library methods from Python. In BM-3~\cite{bm3}, Java methods are called from Python (as host). 

\begin{table}[t]
\centering
\caption{Java-Python apps in GraalVM \label{java-pyhton}}
\vspace{-3mm}
\begin{tabular}{llll}
\hline
Benchmark & Wala-Java & Wala-Python & Conflux \\ \hline
BM-1 & \cmark & \cmark & \cmark \\ 
BM-2 & \cmark & \cmark & \cmark \\
BM-3 & \xmark & \cmark & \xmark \\
\hline
\end{tabular}
\vspace{-1mm}
\end{table}

\toolname~leverages the Wala framework as the pre-analysis, configured to one-callsite-sensitive and field-sensitive for both languages, to obtain the analysis of Java and Python programs. \toolname~successfully analyzed BM-1 and BM-2, and the inter-language function calls are correctly resolved. BM3 uses Java as the guest language and Python as the host. Therefore, Wala-Java failed to analyze the program as it requires a main method, whereas the guest language is used as a library.

\begin{mdframed}[backgroundcolor=gray!20]
\textbf{RQ2:}~\toolname~can be configured to multiple polyglot frameworks. To exemplify,  \toolname~is configured to analyze GraalVM programs, and successfully resolves the bridge interfaces in two of three Java-Python apps.
\end{mdframed}

\section{Related Work}\label{sec:relw}

\paragraph*{Modular polyglot analysis.} Lee~\etal~\cite{Ryu-Semantic-Summary} proposed a modular semantic summary extraction technique for analyzing Java-C/C++ interoperations. Mon\"et~\etal~\cite{Monat-Python-SAS} used abstract interpretation for analyzing Python and C programs. JN-SAF~\cite{weiJNSAF-CCS2018} by Wei~\etal used a function summary based bottom-up dataflow analysis for analyzing Java Native applications. All these works rely on explicit boundaries between the host and guest. Also, these works are tied to particular host and guest languages, while our work is language-agnostic. Buro~\etal proposed an abstract interpretation-based framework~\cite{buro2020abstractmultlingual} for analyzing multilingual programs by combining analysis of while and expression language. In contrast to their work, our work combines languages with more complex operations.

\paragraph*{Monolithic Polyglot Analysis} JET~\cite{tan2011oopsla}, Li and Tan~\cite{tan2009ccs}, Kondoh and Onodera~\cite{Kondoh2008} analyze bugs occurring at Java JNI interfaces. Fourtounis~\etal~\cite{binaryscanning} scanned native binaries to extract the dataflow facts use it for the analysis in Java. Compared to them, our approach is generic and not restricted to the JNI calls. Bae~\etal\cite{hydridDroidICSE2019}, ~\cite{hybridDroid}, Lee~\etal~\cite{adlib2019ryu} proposed techniques to analyze Android WebViews (Java-Javascript) programs. Samhi~\etal~\cite{JuCify2022ICSE} mapped the native code to Soot IR for a unified analysis.  Turcotte~\etal~\cite{gregor2019ecoop} implement a type checker for correctness of types for Lua-C programs. Brucker and Herzberg~\cite{cordova2016esoss} proposed a static analysis to detect security vulnerabilities in Cordova (Java-Javascript) based applications. In contrast to these works on multilingual program analysis, our work defines a modular program analysis combining existing analysis.

\paragraph*{Modular Analysis of monolingual applications}
It is orthogonal to our work, yet these form the backbone of our analysis as pre-analysis. \textsc{JAM}~\cite{nielsen2021issta} by Nielsen~\etal proposed a modular call-graph construction for analysis of \emph{Node.js} applications. Tiwari~\etal~\cite{iifa} used persistent summaries for resolving inter-component communication in Android. \textsc{ModAnalyzer} by Schubert~\etal~\cite{schubert2021ecoop} leverages the pattern used by C++ developers to perform a modular analysis of C++ applications and libraries.
Dilling~\etal~\cite{dilling2011pldi} and Whaley and Rinard~\cite{rinhard1999pointer} proposed compositional pointer analyses for C and Java respectively. Illous~\etal~\cite{rival2020shape} proposed a separation logic based technique for compositional shape analysis.

\section{Conclusion}
In this work, we leveraged existing language analyses for analyzing multilingual programs. We proposed a series of algorithms for the same and demonstrated our approach’s efficacy with a prototype implementation called \toolname. Our analysis neither requires us to remodel the analysis for polyglot applications nor translate the existing semantics of the guest language to the host languages, consequently saving a lot of effort in this process.

\bibliographystyle{ACM-Reference-Format}
\bibliography{RelatedWork}

\newpage

\end{document}